\documentclass[12pt,numbers,sort&compress]{elsarticle}
\usepackage{amsmath}
\usepackage{amssymb}
\usepackage{hyperref}
\usepackage{xcolor}
\usepackage{fvextra}
\DefineVerbatimEnvironment{verbatim}{Verbatim}{
  breaklines=true,
  breakanywhere=true,
  breaksymbolleft={},
  breaksymbolright={},
  breakanywheresymbolpre={},
  breakanywheresymbolpost={}
}
\usepackage{placeins}
\usepackage{graphicx}

\usepackage[a4paper,margin=2.2cm]{geometry}

\makeatletter
\def\ps@pprintTitle{%
  \let\@oddhead\@empty
  \let\@evenhead\@empty
  \let\@oddfoot\@empty
  \let\@evenfoot\@empty}
\makeatother

\hypersetup{
  hypertexnames=false,
  colorlinks=true,
  linkcolor=blue,
  citecolor=blue,
  urlcolor=blue
}

\begin{document}

\begin{frontmatter}

\title{\texttt{PTagent}: An LLM-assisted tool for finite-temperature model construction in cosmological phase transitions}

\author[a,b,c]{Lei Wu}
\ead{leiwu@njnu.edu.cn}
\author[a]{Yang Xiao}
\ead{xiaoyangphy@gmail.com}
\author[a,d]{Jin Min Yang}
\ead{jmyang@itp.ac.cn}
\author[a]{Yang Zhang}
\ead{zhangyang2025@htu.edu.cn}

\address[a]{School of Physics, Henan Normal University, Xinxiang 453007, P. R. China}
\address[b]{Department of Physics and Institute of Theoretical Physics, Nanjing Normal University, Nanjing 210023, P. R. China}
\address[c]{Nanjing Key Laboratory of Particle Physics and Astrophysics, Nanjing 210023, P. R. China}
\address[d]{Institute of Theoretical Physics, Chinese Academy of Sciences, Beijing 100190, P. R. China}

\begin{abstract}
Cosmological phase-transition calculations require well-defined finite-temperature numerical models, whose construction involves assembling model information under consistent theoretical prescriptions. In this work we present \texttt{PTagent}, a large language model (LLM) agent-based framework for constructing such models from SARAH model files, research papers, and DRalgo sources.
The LLM agent interprets and organizes model information
and identifies unresolved physical choices, while dedicated backends perform
the model-dependent calculations.
After review and approval of the resulting physical model specification, the backends generate the corresponding CosmoTransitions-style or PhaseTracer-style numerical potentials.
By separating model interpretation from numerical implementation, PTagent provides an inspectable and reproducible approach to finite-temperature model construction for cosmological phase-transition studies.

\vspace{1em}

\end{abstract}

\end{frontmatter}

\section{Introduction}
\label{sec:introduction}

Cosmological phase transitions arise from the evolution of
the finite-temperature free-energy landscape in the early Universe.
The associated vacuum structure and, for first-order transitions,
the nucleation and expansion of bubbles have important implications
for baryogenesis~\cite{Morrissey:2012db},
stochastic gravitational-wave backgrounds
~\cite{Weir:2017wfa,Caprini:2018mtu,Athron:2023xlk},
reheating and dark matter production or
dilution~\cite{Megevand:2007sv,Baker:2019ndr,Xiao:2022oaq},
primordial black-hole formation~\cite{Lewicki:2023ioy,Flores:2024lng},
and cosmological constraints from vacuum
trapping~\cite{Balazs:2023kuk,Biekotter:2022kgf,Biekotter:2021ysx}.
Quantitative studies of these phenomena involve identifying
the temperature-dependent phases and possible transitions,
calculating bubble nucleation and subsequent thermal evolution,
and, when relevant, bubble-wall dynamics.
A broad range of public numerical tools has been developed
for different stages of these calculations~\cite{Wainwright:2011kj,Basler:2024aaf,Athron:2020sbe,Athron:2024xrh,Matuszak:2026xsz,Brdar:2025hxw,Costa:2025pew,Masoumi:2016wot,Athron:2019nbd,Sato:2019wpo,Guada:2020xnz,Camargo-Molina:2013qva,Ekstedt:2022bff,Ekstedt:2024fyq,vandeVis:2025plm,Ekstedt:2023sqc,Si:2025vdt}.
Despite this extensive numerical infrastructure,
applying these tools to a given particle-physics model
generally requires a well-defined finite-temperature
effective potential, or an equivalent numerical thermal
model, as input.

Constructing such a finite-temperature model is itself a nontrivial part of the calculation.
Even within the conventional perturbative treatment based on the finite-temperature effective potential, one must specify the field conventions, particle spectrum and thermal corrections, together with the renormalization and resummation prescriptions~\cite{Coleman:1973jx,Dolan:1973qd,Parwani:1991gq,Arnold:1992rz}.
These choices define the finite-temperature approximation and can affect the phase structure, tunnelling dynamics, and gravitational-wave predictions~\cite{Athron:2022jyi,Schicho:2022wty}.
In realistic applications, the required information is often distributed across model files, equations, text, supplementary material, and existing code, while some conventions remain implicit.
The practical task is therefore not merely to implement a potential function, but to assemble a complete and internally consistent finite-temperature numerical model.

The perturbative effective-potential approach also carries intrinsic theoretical limitations.
In gauge theories, finite-order calculations can exhibit gauge dependence, infrared sensitivity, and residual dependence on the organization of perturbation theory and thermal resummation~\cite{Nielsen:1975fs,Patel:2011th}.
Dimensional reduction provides a complementary approach in which the hierarchy of thermal scales is treated systematically through a three-dimensional effective field theory~(EFT), with the long-distance theory amenable to higher-order perturbative or nonperturbative analysis~\cite{Kajantie:1995dw,Ekstedt:2022bff}.
These considerations further emphasize that the finite-temperature theory and approximation used in a phase-transition calculation must be specified explicitly.

Large language model (LLM) agents are increasingly
being applied to high-energy physics research
~\cite{Qiu:2026iby,Menzo:2026xat,Xiao:2026gal,Guo:2026ifi},
and their ability to interpret and organize scientific
information makes them well suited to the
model-construction task discussed above.
Recent studies have begun to apply agentic workflows to cosmological phase-transition calculations.
DarkAgents presents a multi-agent system for theoretical astroparticle physics in which LLMs support reasoning, code generation, workflow construction, and assumption auditing, while deterministic and tested human-written code performs the underlying physics calculations~\cite{Lucente:2026kgh}.
Its phase-transition implementation, DarkAgent-PT, uses a dedicated semi-analytic finite-temperature backend for classically scale-invariant models with a single scalar background field.
LeWRON presents an agentic framework that orchestrates the electroweak phase-transition analysis pipeline starting from an input Lagrangian~\cite{Wang:2026jjn}.
It constructs model-specific symbolic and numerical tools, uses auditor agents to examine intermediate analytic results, and employs the resulting code for thermal-history calculations and further numerical exploration.
These studies illustrate how LLM agents can organize model-dependent information and multi-stage workflows in cosmological phase-transition calculations.

In this work, we present \texttt{PTagent}, an
LLM agent-based framework that focuses on the construction
of model-dependent finite-temperature numerical potentials.
Rather than attempting to automate the entire
phase-transition analysis workflow, PTagent addresses
the model-construction stage and provides numerical
potentials that can be used by established software
for subsequent phase-transition calculations.
PTagent is implemented as an agent skill that can
be integrated into compatible LLM-agent environments.
This design allows model construction to become part
of interactive research workflows in which LLM agents
increasingly assist scientific work.
The agent interprets and organizes model information
and identifies unresolved physical choices for 
review, while dedicated backends perform the
model-dependent calculations and generate
CosmoTransitions-style or PhaseTracer-style potentials
from the approved model specification.

The remainder of this paper is organized as follows.
Section~\ref{sec:quick-start} provides a quick start guide to installing
and using \texttt{PTagent} through representative examples.
Section~\ref{sec:framework} describes the construction of finite-temperature
numerical models, including the required physical ingredients,
model specification, and backend implementation and validation.
Section~\ref{sec:applications} presents three applications based on SARAH model files,
a research paper, and a DRalgo source.
Finally, Section~\ref{sec:conclusion} summarizes our conclusions.

\section{Quick Start}
\label{sec:quick-start}

PTagent is available at
\url{https://github.com/Contract-Mediated-Agent/PTagent}.
It is distributed as an agent skill with backends.
The repository can be copied or cloned into the skill directory of a compatible
agent environment. Alternatively, the repository address can be given directly
to an agent that supports skill installation. For example, in OpenAI Codex or DeepSeek Harness one may ask
\begin{verbatim}
Install the PTagent skill from
https://github.com/Contract-Mediated-Agent/PTagent
\end{verbatim}
Then the skill can be invoked as \texttt{\$ptagent} in Codex or \texttt{\textbackslash ptagent} in DeepSeek Harness.
PTagent checks the local environment on first use and reports missing
dependencies required by the selected workflow. The examples below use
SARAH 4.14.5~\cite{Staub:2013tta},
CosmoTransitions 2.0.7 for the first example, and PhaseTracer 2.2.2 for the second.

The following examples illustrate two typical ways of starting a PTagent
calculation. The detailed inputs in these examples are provided for
reproducibility, but they need not all be specified in the
initial request. PTagent can interactively clarify unresolved
physical choices, and an existing model contract can be
modified and reused without repeating the source extraction.

\subsection{Begin with a SARAH model}

As a first example, we consider the type-I two-Higgs-doublet model~(THDM) distributed
with SARAH. Replacing the path below by the location of the local SARAH
installation, the calculation can be started with
\begin{verbatim}[commandchars=\\\{\}]
$ptagent Study the EWPT of /path/to/SARAH/Models/THDM with CosmoTransitions.
\end{verbatim}

PTagent first imports the tree potential, parameter relations, and field-dependent
mass matrices from the SARAH export. The export used here does not contain the
thermal self-energies. These are obtained by a separate symbolic calculation from
the scalar and Yukawa interactions and gauge representations in the original
SARAH model files, as described in Sec.~\ref{sec:source-information}.
The numerical calculation uses the default example parameters supplied with the
SARAH model. After the derived thermal inputs and physical prescriptions are
reviewed and approved, the dedicated backend generates the CosmoTransitions
implementation and runs the phase-transition calculation.

The thermal completion and checks of the constructed model are described
in Sec.~\ref{sec:sarah-thdm}.

\subsection{Begin with a paper}

PTagent can also construct a finite-temperature model directly from a paper.
As an example, consider the \(Z_2\)-symmetric real-singlet extension of the
Standard Model (xSM) studied in
Ref.~\cite{Xiao:2022oaq}. The benchmark calculation used here is started with

\begin{verbatim}[commandchars=\\\{\}]
$ptagent Study the xSM model in arXiv:2207.14519 and perform the calculation with PhaseTracer. Use the input parameters mS = 572, lambda_hs = 9.7, lambda_s = 0.96, with vh=246 and mh=125. Include scalars, W/Z longitudinal and transverse modes, and the top quark; use num_boson_dof=29 and num_fermion_dof=90. Use the Parwani method. Do not include Goldstone bosons or photons in the mass lists. The Hessian matrix needs to be corrected.
\end{verbatim}

The agent connects the model information and thermal prescription in the paper
with the explicit choices supplied by the user. Any remaining physical
ambiguities that affect the generated model are returned for review. After the
model specification is approved, the dedicated backend constructs the
PhaseTracer implementation and runs the transition calculation.

The adopted prescriptions and checks of the numerical potential are
described in Sec.~\ref{sec:xsm}.
Local PDF, TeX, or Markdown sources can be used in the same way as an arXiv
identifier.

\section{Model construction with PTagent}
\label{sec:framework}

\begin{figure}[t]
\centering
\includegraphics[width=\textwidth]{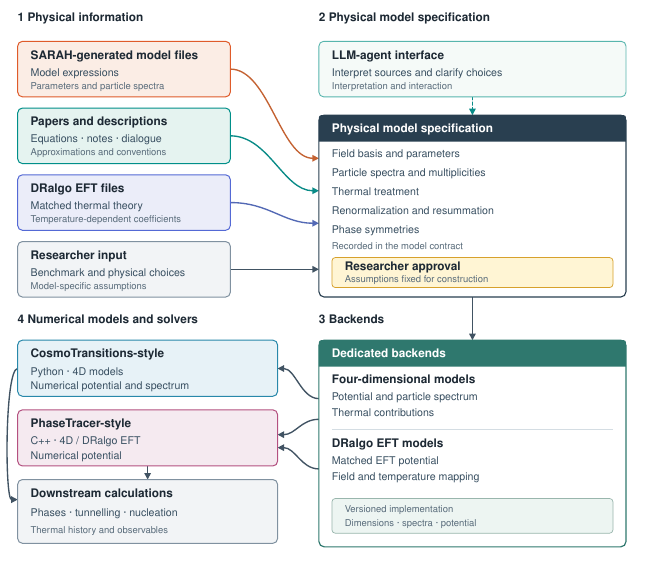}
\caption{Architecture of PTagent. Source information and user choices define the physical model specification. The LLM agent assists interpretation and clarification, while dedicated backends construct the numerical potentials. }
\label{fig:ptagent-workflow}
\end{figure}

The primary purpose of PTagent is to construct the finite-temperature
free energy of a given particle-physics model in a form suitable for
numerical phase-transition calculations. As illustrated in Fig.~\ref{fig:ptagent-workflow},
PTagent combines an LLM-agent interface with dedicated computational
backends. The agent interprets model information, organizes the required
physical ingredients, and identifies unresolved choices for
review. Once the resulting model specification is approved, the backends
construct the corresponding finite-temperature numerical model for
subsequent phase-transition calculations.

The inputs to this workflow are discussed in the following subsections, while the resulting numerical potentials are provided in two formats, CosmoTransitions-style and PhaseTracer-style, to facilitate integration with existing
phase-transition calculation frameworks. The CosmoTransitions-style
output follows the Python potential interface of CosmoTransitions 2.0.7~\cite{Wainwright:2011kj},
which is also used by ELENA v1.0~\cite{Costa:2025pew}
and forms the basis of TransitionListener v2.0.1~\cite{Matuszak:2026xsz}.
The PhaseTracer-style output implements the C++ potential interface
of PhaseTracer~\cite{Athron:2020sbe,Athron:2024xrh}, which is also used by TransitionSolver~\cite{Athron:2022mmm}.
These two formats allow PTagent-generated models to be used in
downstream calculations of phase structure, bubble nucleation,
thermal history, and gravitational-wave observables.

\subsection{Finite-temperature effective potentials}
\label{sec:finite-temperature-model}

PTagent supports both four-dimensional perturbative effective
potentials and dimensionally reduced three-dimensional effective
theories. In the four-dimensional approach, quantum and thermal
effects are incorporated as corrections to the tree-level
scalar potential. The finite-temperature effective potential
can be written as
\begin{equation}
    V_{\mathrm{eff}}(\varphi,T)
    = V_0 + V_{\mathrm{CW}} + V_{\mathrm{CT}}
    + V_T + V_{\mathrm{ring}},
    \label{eq:veff-decomposition}
\end{equation}
where $\varphi$ denotes the scalar background fields.
The quantum and thermal corrections depend on the
adopted theoretical approximations and prescriptions,
including gauge fixing, renormalization, and thermal
resummation. To illustrate how these contributions are constructed,
we consider a commonly used perturbative formulation
and discuss the relevant choices where necessary.
This formulation is illustrative rather than restrictive,
as \texttt{PTagent} can accommodate other theoretical
prescriptions.

The tree-level potential $V_0(\varphi)$ is obtained by
evaluating the scalar potential of the Lagrangian at the
chosen background-field configuration. Quantum fluctuations
around this background give rise to the zero-temperature
one-loop correction, which is determined by the field-dependent
particle spectrum. For example, in Landau gauge and the
$\overline{\mathrm{MS}}$ scheme, the real part of the
Coleman--Weinberg~(CW) potential~\cite{Coleman:1973jx} is given by
\begin{equation}
    V_{\mathrm{CW}}(\varphi)
    = \sum_i \frac{(-1)^{F_i} n_i}{64\pi^2}
    m_i^4(\varphi)
    \left[
        \ln\frac{|m_i^2(\varphi)|}{Q^2}-c_i
    \right],
    \label{eq:CW}
\end{equation}
where $m_i^2(\varphi)$ are the field-dependent squared-mass
eigenvalues, $n_i$ denotes the corresponding degrees of freedom~(d.o.f.),
$F_i=0$ for bosons and $F_i=1$ for fermions, and $Q$ is
the renormalization scale. The constants are $c_i=3/2$
for scalars and fermions and $c_i=5/6$ for gauge bosons.
The counterterm potential $V_{\mathrm{CT}}$, when included,
is determined by the adopted renormalization conditions.
Other prescriptions, such as on-shell-like (OS-like) treatments,
may combine the one-loop and counterterm contributions
into a single expression.

The field-dependent masses entering the loop corrections
are obtained by expanding the Lagrangian around the chosen
scalar background. For the scalar sector, the squared-mass
matrix is given by
\begin{equation}
    M^2_{S,ab}
    =
    \left.
    \frac{\partial^2 V_0(\phi)}
    {\partial\phi_a\partial\phi_b}
    \right|_{\phi=\phi(\varphi)},
    \label{eq:scalar-mass-matrix}
\end{equation}
where $\phi$ contains all canonically normalized real scalar
components and $\phi(\varphi)$ specifies the chosen background
configuration. The derivatives include charged and Goldstone
fluctuations even when their background values vanish.
The gauge-boson squared-mass matrices follow from the
covariant scalar kinetic terms, while the fermion squared
masses are obtained from the eigenvalues of
$M_F^\dagger M_F$, where $M_F$ is derived from the
fermion mass and Yukawa interactions.

The field-dependent particle spectrum also determines the
finite-temperature contribution to the effective potential.
In thermal equilibrium at vanishing chemical potentials,
the one-loop thermal correction takes the form
\begin{equation}
    V_T(\varphi,T)
    = \frac{T^4}{2\pi^2}
    \operatorname{Re}
    \left[
        \sum_b n_b J_B\left(\frac{m_b^2(\varphi)}{T^2}\right)
        -
        \sum_f n_f J_F\left(\frac{m_f^2(\varphi)}{T^2}\right)
    \right],
    \label{eq:thermal}
\end{equation}
where the sums run over bosonic and fermionic degrees of freedom,
respectively. The thermal functions~\cite{Dolan:1973qd} are defined by
\begin{equation}
    J_{B,F}(y)
    = \int_0^\infty dx\,x^2
    \ln\left(1 \mp e^{-\sqrt{x^2+y}}\right),
    \label{eq:thermal-functions}
\end{equation}
with the upper and lower signs corresponding to bosons and
fermions, respectively.
The particle species included in the thermal sums must be
specified consistently with the adopted approximation.
Additional relativistic degrees of freedom that are not
included explicitly in the field-dependent spectrum may
contribute to the field-independent radiation part of
the free energy.

Although the one-loop thermal potential incorporates thermal
fluctuations, perturbative calculations can suffer from
infrared sensitivity associated with bosonic zero Matsubara
modes. This motivates the resummation of thermal corrections,
which introduces screening effects through temperature-dependent
bosonic self-energies. Unlike the field-dependent masses
obtained from the tree-level Lagrangian, these thermal
self-energies arise from interactions with the thermal
medium. Their leading high-temperature contributions
can be calculated from the scalar, gauge, and Yukawa
interactions of the model~\cite{Basler:2018cwe}.

For scalar fields, the leading one-loop high-temperature
self-energies can be expressed in terms of the scalar quartic
couplings, gauge interactions, and Yukawa couplings as~\cite{Basler:2018cwe}
\begin{equation}
    \left(\Pi_S\right)_{ab}(T)
    = T^2 \left[
        \frac{1}{24}\sum_c \lambda_{abcc}
        + \frac{G_{ab}}{4}
        + \frac{Y_{ab}}{12}
    \right],
    \label{eq:scalar_thermal_mass}
\end{equation}
where
$\lambda_{abcd} =
\partial^4 V_0/
(\partial\phi_a\partial\phi_b\partial\phi_c\partial\phi_d)$
is the scalar quartic tensor defined in the full real-scalar
basis. The first term therefore includes contributions
from all scalar components, including those without
nonzero background values.
The matrix $G_{ab}$ represents the gauge contribution
determined by the coupling-weighted quadratic Casimir
operators acting on the scalar fields, while
$Y_{ab}=\operatorname{Re}\operatorname{Tr}
(Y^aY^{b\dagger})$ denotes the Yukawa contraction.
Here the Yukawa matrices are defined in the
two-component fermion convention through
$\mathcal{L}_Y \supset
-\frac{1}{2}(Y^a)_{IJ}\phi_a\psi_I\psi_J+\mathrm{h.c.}$,
where $I,J$ include all internal and generation indices.

Gauge interactions also generate thermal screening
for the longitudinal components of gauge bosons.
At leading order, their Debye self-energies can be
written as
\begin{equation}
    \left(\Pi_L\right)_{AB}(T)
    = T^2 \left[
        \frac{A_{AB}}{3}
        + \frac{S_{AB}}{3}
        + \frac{F_{AB}}{6}
    \right],
    \label{eq:gauge_thermal_mass}
\end{equation}
where $A_{AB}$, $S_{AB}$, and $F_{AB}$ are the
coupling-weighted representation traces over
adjoint gauge fields, complex scalars, and
Weyl fermions, respectively.
For a simple non-Abelian gauge group, the adjoint
contribution is $A_{AB}=g^2 C_A\delta_{AB}$,
where $C_A$ is the adjoint quadratic Casimir.
The scalar and fermion traces account for their
gauge representations and multiplicities in the
thermal plasma~\cite{Ekstedt:2022bff}.

The resulting self-energies modify the bosonic
mass matrices according to
\begin{equation}
    \mathcal{M}_b^2(\varphi,T)
    =
    \mathcal{M}_{b,0}^2(\varphi)
    + \Pi_b(T).
    \label{eq:resummed_mass}
\end{equation}
The thermally corrected mass eigenvalues are obtained
by diagonalizing these matrices after adding the
self-energies. This is particularly important when
different scalar or gauge fields mix in the presence
of background fields. For gauge bosons, the leading
Debye corrections apply only to the longitudinal
polarizations. The expressions above are derived
in the high-temperature approximation and therefore
do not account for the Boltzmann suppression of
particles with masses much larger than the temperature.

The incorporation of these thermal corrections into
the effective potential depends on the resummation
prescription. In the Arnold--Espinosa approach~\cite{Arnold:1992rz},
the unresummed masses are retained in
$V_{\mathrm{CW}}$ and $V_T$, while the screened
bosonic zero modes contribute an additional ring term,
\begin{equation}
    V_{\mathrm{ring}}(\varphi,T)
    =
    -\frac{T}{12\pi}
    \sum_{b\in\mathrm{screened\ modes}}
    n_b^{\mathrm{ring}}
    \operatorname{Re}
    \left[
        \left(M_b^2(\varphi,T)\right)^{3/2}
        -
        \left(m_b^2(\varphi)\right)^{3/2}
    \right],
    \label{eq:ring}
\end{equation}
where $M_b^2(\varphi,T)$ and $m_b^2(\varphi)$
denote the thermally corrected and unresummed
squared-mass eigenvalues, respectively.
The multiplicities $n_b^{\mathrm{ring}}$ count
the screened scalar and longitudinal gauge modes,
excluding transverse gauge polarizations.
Alternatively, in the Parwani prescription~\cite{Parwani:1991gq}, the thermally corrected bosonic masses enter
both $V_{\mathrm{CW}}$ and $V_T$, without a separate
ring contribution. These prescriptions differ in
their treatment of higher-order thermal corrections
and can lead to different numerical predictions.

The treatment of negative squared masses introduces
another prescription dependence. In the real-part
prescription considered here, the thermal functions
and cubic mass terms are analytically continued
to negative squared masses, and their real parts
are retained~\cite{Basler:2018cwe}. 
Such choices, together with the treatment of
Goldstone bosons and the particle content included
in the thermal sums, must be specified consistently
when constructing the finite-temperature potential.

Beyond the conventional four-dimensional perturbative approach,
dimensional reduction~\cite{Kajantie:1995dw} provides an alternative framework for
studying thermal phase transitions. By systematically integrating
out the higher-energy thermal modes, this approach constructs
a three-dimensional effective field theory that describes
the long-distance dynamics of the system.
The parameters of the resulting theory are determined through
matching to the underlying four-dimensional model.
The three-dimensional effective potential can then be studied
perturbatively or, where appropriate, using nonperturbative
methods.

In this approach, the numerical backend uses a matched
three-dimensional effective theory, supplied directly or obtained
by running the provided DRalgo model program~\cite{Ekstedt:2022bff}. The matched theory provides the effective
potential and its temperature-dependent parameters.
To interface with numerical phase-transition calculations,
the three-dimensional potential is converted into
its contribution to the four-dimensional free-energy density.
At leading order in the field normalization, the
conversion is given by
\begin{equation}
    \varphi_{i,\mathrm{3d}}
    = \frac{\varphi_i}{\sqrt{T}},
    \qquad
    V_{\mathrm{4d}}(\varphi,T)
    = T V_{\mathrm{3d}}(\varphi_{\mathrm{3d}},T).
    \label{eq:3deft_mapping}
\end{equation}
Here $\varphi_{i,\mathrm{3d}}$ denotes the scalar fields
of the three-dimensional effective theory, while
$\varphi_i$ represents the corresponding four-dimensional
background fields. Higher-order matching may introduce
additional field-normalization factors, which must be
taken consistently from the upstream calculation~\cite{Kajantie:1995dw}.

Discrete symmetries of the effective potential also play an
important role in identifying phases and transitions.
Although symmetry-related minima have identical free energies,
transitions between them and other vacua may correspond to
inequivalent tunnelling paths.
CosmoTransitions allows users to restrict the phase search
through a user-defined filtering condition~\cite{Wainwright:2011kj}, whereas PhaseTracer
supports explicit discrete symmetry transformations to identify
equivalent phases while retaining inequivalent transitions~\cite{Athron:2020sbe,Athron:2024xrh}.
The relevant symmetries and their treatment in the target
framework must therefore be specified consistently.

Together, these ingredients and prescriptions define the
finite-temperature numerical model to be constructed.
Some can be determined from the underlying particle-physics
model, while others require additional theoretical choices.
The following section describes how PTagent obtains,
organizes, and verifies this information from different
sources.

\subsection{From source information to a reviewed model}
\label{sec:source-information}

How PTagent constructs a finite-temperature numerical model
from the physical ingredients discussed above depends
on the information available from the input source.
For models implemented in SARAH, the Vevacious++ export
provides the tree-level potential, field-dependent scalar,
gauge-boson and fermion squared-mass matrices, particle
multiplicities, and parameter relations.
Additional information about scalar and Yukawa
interactions and gauge representations is obtained from
the original SARAH model files, while numerical inputs
can be supplied through an SLHA format or a user-specified
benchmark.

However, the zero-temperature export generally does not
contain all the thermal self-energies required for resummation.
To obtain these corrections, the agent identifies and organizes
the relevant scalar interactions, Yukawa couplings, and gauge
representations from the original model files.
Once these interactions and the thermal particle content are
specified, symbolic routines evaluate the required tensor
contractions, and the backend constructs the thermal
self-energies using the expressions introduced in Sec.~\ref{sec:finite-temperature-model}.
The scalar calculation requires interactions
involving the complete scalar multiplets, including components
without background values.
This is illustrated by the THDM in
Sec.~\ref{sec:sarah-thdm}. Missing interactions or unspecified thermal
particle content are returned for user clarification.

When starting from a paper or description,
the main challenge is to interpret and organize model
information distributed across equations, text, and
supplementary material. The agent identifies the tree-level
potential, field-dependent particle masses, and quantum
and thermal corrections already specified in the source.
Distinguishing these contributions is essential to avoid
double counting or incorrectly treating a thermal
self-energy as a complete thermal correction.
The renormalization, resummation, and other physical
prescriptions are taken from the source when available,
while unspecified choices are returned to the user.
Possible inconsistencies or typographical errors are also
identified for review, as illustrated by the correction
to the scalar mass matrix in the xSM example of Sec.~\ref{sec:xsm}.

The DRalgo route differs from the preceding cases because
the finite-temperature theory is provided through
dimensional reduction. PTagent uses the supplied DRalgo
model program and its matching results to construct
the corresponding numerical potential, following the
conversion in Eq.~\eqref{eq:3deft_mapping}.
The temperature dependence of the matched coefficients,
field normalization, and selected matching and
effective-potential orders are preserved without
introducing additional matching approximations.

Regardless of the input source, the resulting physical
information is assembled into a human-readable model
contract that records the model expressions, their
sources, numerical inputs, and adopted prescriptions.
For choices not determined by the available sources,
the agent explains the ambiguities and their physical
implications, and provides suggestions for
review when appropriate. The user's decisions
are then incorporated into the model specification.
Once approved, this specification defines the
finite-temperature model to be constructed by the
dedicated backends, as described in the following section.

\subsection{Backend implementation and validation}

Once the physical model specification has been approved,
PTagent uses dedicated backends to construct the corresponding
numerical effective potentials. For four-dimensional models,
the backends implement the specified particle spectra,
quantum and thermal corrections, and resummation prescriptions
in CosmoTransitions-style or PhaseTracer-style formats.
For dimensionally reduced models, the matched three-dimensional
potential is converted into a PhaseTracer-style implementation
using the field normalization and temperature dependence
given in Eq.~\eqref{eq:3deft_mapping}. In both cases, the backends follow
the approved model specification without introducing
additional physical assumptions.
Backend generation is blocked if essential physical
information or prescriptions remain unresolved.
For four-dimensional models, recorded hashes link the generated
implementations to the approved model specification, allowing
outdated implementations to be identified when the specification
is modified.

The generated models are validated at several levels,
beginning with dimensional consistency, field conventions,
and particle spectra, followed by checks of thermal self-energies,
potential values, and numerical interfaces.
For SARAH inputs, the derived scalar thermal self-energies
are independently checked against the Hessian of the
leading $T^2$ thermal contribution obtained from the
unresummed particle spectrum.
Pointwise comparisons between different output formats and
independent evaluation of the supplied EFT expressions further
test the numerical implementation.

For fixed model inputs, physical prescriptions, and
backend version, the resulting numerical potential
is reproducible. We have tested PTagent with
GPT-6 Astra, GPT-5.6 Sol, and DeepSeek-V4.1-Flash using the examples presented in this work, obtaining consistent effective potentials.
The following section illustrates three model-construction
routes and checks of the resulting numerical potentials.

\FloatBarrier
\section{Model construction and validation}
\label{sec:applications}

The construction of a finite-temperature numerical model depends
on how the underlying physical information is supplied.
We illustrate three representative cases, namely a THDM exported from SARAH, an xSM model
described in a paper, and a matched three-dimensional
effective theory obtained with DRalgo.

\subsection{Model construction from SARAH}
\label{sec:sarah-thdm}

For the first example in Sec.~2, PTagent starts from the
type-I THDM model files in SARAH, generates the
Vevacious++ export, and imports the resulting files.
These provide the tree-level potential, field-dependent
scalar, gauge-boson and fermion mass matrices,
particle multiplicities, and parameter relations.
The original SARAH model definitions also provide
the scalar interactions, gauge representations, and
Yukawa couplings needed for thermal corrections.
While the exported information is sufficient to construct
the unresummed one-loop effective potential, the thermal
self-energies required for resummation must be obtained
separately.

\begin{figure}[!t]
\centering
\includegraphics[width=\textwidth]{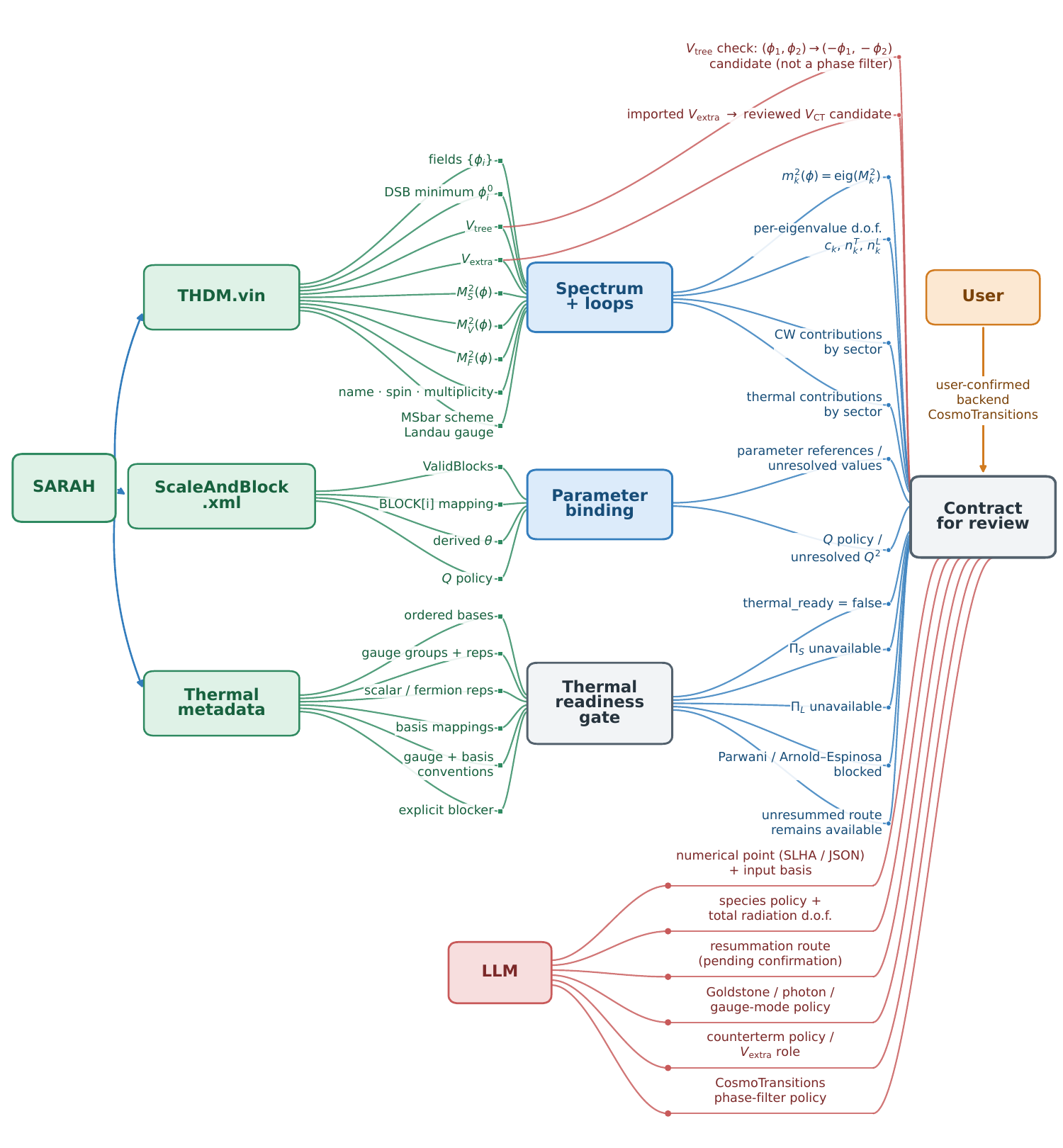}
\caption{Provenance graph generated by PTagent for the SARAH THDM contract
at the initial review stage. Green denotes information imported from SARAH
exports and companion metadata, blue denotes deterministic processing,
orange denotes user-confirmed choices, and pale red denotes proposals
awaiting confirmation. }
\label{fig:thdm-contract-provenance}
\end{figure}

To calculate these thermal self-energies, PTagent uses
the interactions obtained from the original SARAH model.
In particular, the scalar thermal corrections require
the full gauge-invariant scalar potential.
The symbolic calculation therefore expands the potential
in the eight canonically normalized real components of
the two Higgs doublets and evaluates the quartic
contractions according to Eq.~\eqref{eq:scalar_thermal_mass}.
The gauge contributions are determined by the quadratic
Casimir operators, while the Yukawa contributions are
obtained from the fermion mass matrices and the
corresponding Yukawa assignments.
In the type-I model, all quarks and charged leptons
couple to $H_2$, so their Yukawa contributions enter
only the thermal self-energy of this doublet.
The longitudinal gauge-boson self-energies are similarly
derived from the gauge and matter representations
following Eq.~\eqref{eq:gauge_thermal_mass}.

The quartic part of the scalar potential is
\begin{equation*}
\begin{aligned}
V_0 \supset{}&
\Lambda_1(H_1^\dagger H_1)^2
+\Lambda_2(H_2^\dagger H_2)^2
+\Lambda_3(H_1^\dagger H_1)(H_2^\dagger H_2)\\
&+\Lambda_4|H_1^\dagger H_2|^2
+\left[\frac{\Lambda_5}{2}(H_1^\dagger H_2)^2+\mathrm{h.c.}\right].
\end{aligned}
\end{equation*}
In this convention, the scalar thermal self-energies are
\begin{equation}
\begin{aligned}
\Pi_{H_i}(T)
&= T^2 \left[
\frac{3g^2+g'^2}{16}
+\frac{\Lambda_i}{2}
+\frac{2\Lambda_3+\Lambda_4}{12}
+\frac{\delta_{i2}Y_2}{12}
\right],\\
Y_2
&= \operatorname{Tr}\left(
3Y_u^\dagger Y_u
+3Y_d^\dagger Y_d
+Y_e^\dagger Y_e
\right).
\end{aligned}
\end{equation}
Here $i=1,2$, and $g$ and $g'$ denote the electroweak
gauge couplings. The same thermal correction applies
to all four real components of each Higgs doublet,
while the contribution from $\Lambda_5$ cancels
in the quartic contraction.
For the longitudinal gauge bosons, the thermal
self-energies are given by
$T^2\operatorname{diag}(2g'^2,2g^2,2g^2,2g^2)$
in the $(B,W^3,W^1,W^2)$ basis.
These corrections are added to the corresponding
field-dependent mass matrices before diagonalization
to obtain the thermally corrected particle spectrum.

To verify the consistency of these thermal corrections
with the original SARAH model, PTagent first checks
that the Hessian of the reconstructed full scalar potential
reproduces the scalar mass matrices exported by SARAH
when evaluated on the neutral background.
As an independent check, the leading high-temperature
expansion of the unresummed particle spectrum is used
to verify the scalar thermal coefficients obtained from
the interaction-based calculation.
These comparisons test the consistency of the field
conventions and the derived thermal self-energies.

After completing these calculations and consistency checks,
PTagent incorporates the derived thermal corrections into
the model contract, which is presented together with a
provenance graph for review.
Figure~\ref{fig:thdm-contract-provenance} shows an initial snapshot
of this process, where the missing thermal information
is explicitly identified.
The graph allows users to trace the sources of
the model information, distinguish imported and derived
quantities, and examine the physical choices that
remain to be confirmed.
Once these choices have been resolved and the completed
contract is approved, PTagent proceeds to construct
the corresponding numerical model.

For this example, the approved specification adopts
the default SARAH benchmark parameters, Landau gauge,
and Parwani resummation with $Q=v_h=246~\mathrm{GeV}$
and no finite counterterms.
The full scalar spectrum, including Goldstone bosons,
and both screened neutral longitudinal gauge modes
are retained.
The backend then generates a CosmoTransitions-style
implementation, which is loaded into CosmoTransitions
for numerical evaluation and phase-transition calculations.

\subsection{Model construction from a paper}
\label{sec:xsm}

Unlike the structured model information provided by SARAH,
the ingredients required to construct a finite-temperature
effective potential from a paper are often distributed
across different equations and discussions.
To illustrate this route, we consider the xSM studied in Ref.~\cite{Xiao:2022oaq}.
With $H^\dagger H=h^2/2$ and $S=s$, the tree-level
scalar potential is
\begin{equation}
    V_0(h,s)
    = -\frac{\mu_h^2}{2}h^2
      +\frac{\lambda_h}{4}h^4
      -\frac{\mu_s^2}{2}s^2
      +\frac{\lambda_s}{4}s^4
      +\frac{\lambda_{hs}}{4}h^2s^2.
\end{equation}
The corresponding finite-temperature model also requires
the particle spectrum, loop corrections, thermal
self-energies, and the prescriptions adopted in the
original calculation. PTagent identifies these ingredients
from the paper and organizes them into a consistent
model specification.

An important part of this process is to distinguish
the physical contributions already included in the
published expressions. In this example, the paper
adopts an on-shell-like one-loop potential that
incorporates finite counterterms, so no additional
counterterm contribution should be introduced.
The finite-temperature calculation employs Parwani
resummation.
PTagent also checks the field-dependent scalar
mass matrix, as requested by the user.
Taking the second derivatives of the tree-level
potential gives
\begin{equation}
\begin{aligned}
    \frac{\partial^2 V_0}{\partial h^2}
    &= -\mu_h^2+3\lambda_h h^2
       +\frac{\lambda_{hs}}{2}s^2,\\
    \frac{\partial^2 V_0}{\partial h\partial s}
    &= \lambda_{hs}hs.
\end{aligned}
\end{equation}
Comparison with the expressions in the source
reveals inconsistent powers of the scalar fields.
The corrected Hessian is therefore adopted after review, ensuring that the scalar spectrum
is consistent with the specified tree-level potential.

\begin{figure}[t]
\centering
\includegraphics[width=\textwidth]{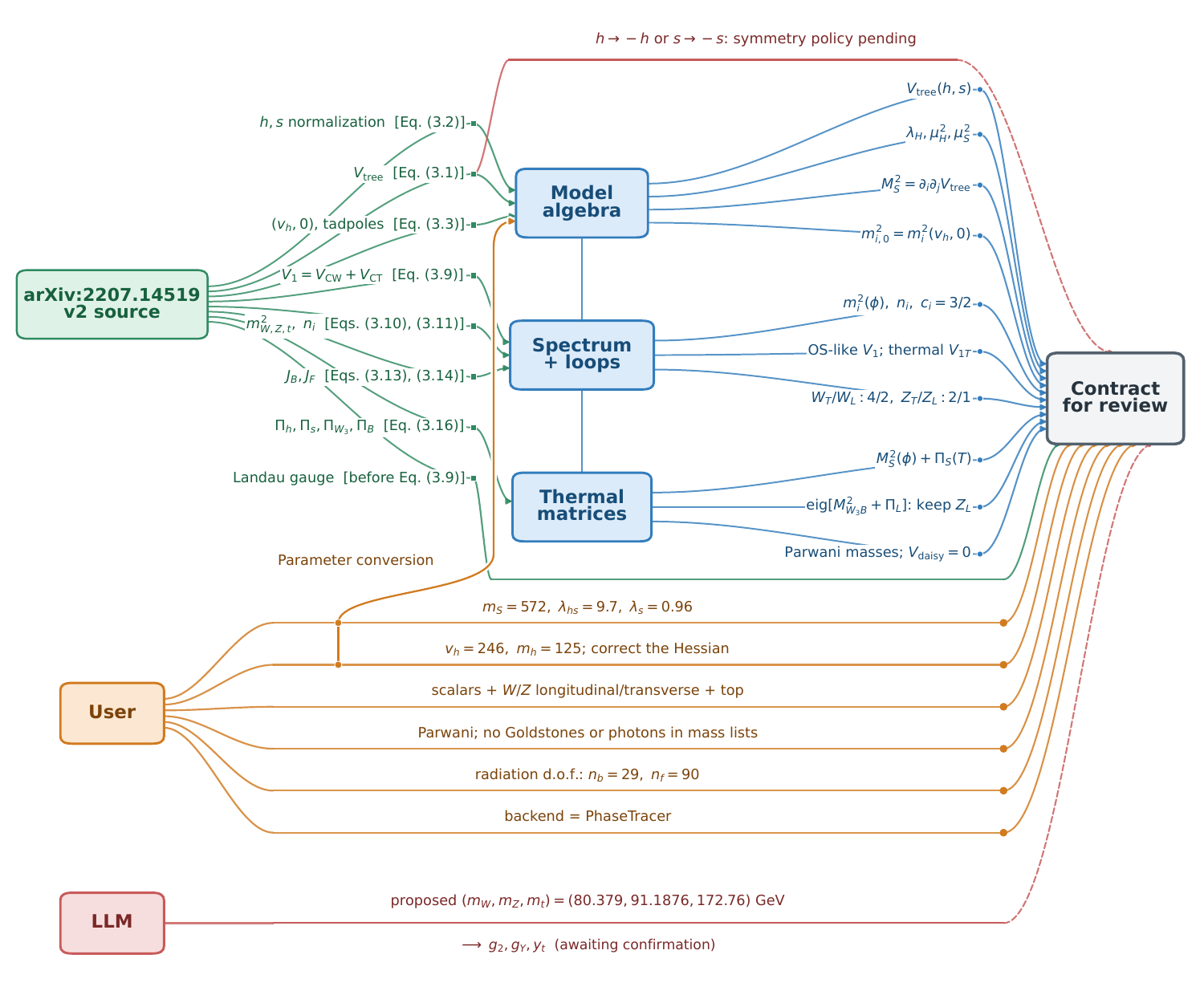}
\caption{Provenance graph for the xSM contract before user approval, using
the same format and color conventions as Fig.~\ref{fig:thdm-contract-provenance}.}
\label{fig:xsm-contract-provenance}
\end{figure}

Once the model information has been assembled,
PTagent presents the contract together with a
provenance graph for review, as shown
in Fig.~\ref{fig:xsm-contract-provenance}.
The figure illustrates how expressions extracted
from the paper, derived quantities, and user
inputs are distinguished from proposals awaiting
confirmation.
For example, the top-quark mass initially proposed
by the agent differs from the value subsequently
adopted in the approved benchmark.
This allows the user to examine the origin
of the physical inputs and resolve ambiguities
before the numerical model is generated.

After approval, the backend constructs the corresponding
PhaseTracer-style finite-temperature potential.
A CosmoTransitions-style implementation is also
generated to provide an independent numerical check.
Both implementations use the same particle content,
thermal corrections, and resummation prescription.
Comparing the effective potentials at 20
field-temperature points gives normalized differences
below $3\times10^{-8}$, which  verifies the numerical consistency
of the two implementations under the approved
prescription.

\subsection{Model construction from a dimensionally reduced EFT}
\label{sec:3deft-application}

We illustrate the three-dimensional EFT workflow using
the xSM implemented in the DRalgo source file
\texttt{xsm\_test.m}, provided in the PTagent repository.
This file is adapted from the official DRalgo xSM example~\cite{Ekstedt:2022bff}
to describe a specific $\mathbb{Z}_2$-symmetric model
and a well-defined dimensional-reduction setup.
In particular, the singlet interactions that are odd
under the $\mathbb{Z}_2$ symmetry are removed, and
the original tutorial's multiple reduction paths are
restricted to a single setup that retains both the Higgs
and singlet scalar fields in the three-dimensional theory.

The adapted file also specifies the Standard Model
hypercharges and employs a consistent mass-basis
treatment, including the rotation of interaction tensors
required for higher-order effective-potential calculations.
Starting from this DRalgo source, PTagent identifies
the relevant model information and invokes DRalgo
to obtain the matched three-dimensional effective theory.
The calculation can be initiated with the following instruction.

\begin{verbatim}[commandchars=\\\{\}]
$ptagent Use xsm_test.m to generate a PhaseTracer model with the default scale policy. Set inputs: mS = 572, lambda_hs = 9.7, lambda_s = 0.96.
\end{verbatim}

From the supplied source, PTagent identifies the field
content, interactions, and dimensional-reduction
prescriptions. It then executes the DRalgo calculation
to obtain the matched parameters and effective potential,
including their temperature dependence.
The specified scale choices and the perturbative orders
defined in the source are retained throughout this process.
The resulting expressions and numerical inputs are
assembled into a model contract for review.
After approval, the backend constructs a PhaseTracer-style
potential using the field normalization and free-energy
conversion given in Eq.~\eqref{eq:3deft_mapping}. The specified discrete
symmetry and low-temperature phase guess are also
incorporated into the generated PhaseTracer implementation.

The generated C++ potential is compared with the
original DRalgo expressions at several representative
field-temperature points, using the same matched
three-dimensional EFT parameters.
The leading-order, next-to-leading-order, and
next-to-next-to-leading-order contributions, as well
as their sum, agree within numerical precision.
These comparisons verify the consistent translation
of the DRalgo effective potential into the
PhaseTracer-style numerical implementation.

\section{Conclusions}
\label{sec:conclusion}

In this work, we presented \texttt{PTagent}, an agent-based framework
for constructing finite-temperature numerical models from
SARAH model files, papers, and DRalgo effective theories.
An LLM agent interprets and organizes the physical information
and identifies unresolved choices for review,
while dedicated backends construct CosmoTransitions-style
and PhaseTracer-style potentials from the approved model
specification. Applications to different models, together
with independent consistency checks, demonstrate the
capabilities of PTagent in finite-temperature model
construction. By making physical assumptions traceable
and numerical implementations independently testable,
PTagent provides a practical and reproducible approach
to preparing particle-physics models for cosmological
phase-transition calculations.

\section*{Acknowledgements}

OpenAI Codex and DeepSeek Harness were used for code assistance and numerical-check preparation.
L. W. thanks Henan Normal University for its hospitality and support during his visit, where part of this work was completed.
This work was supported by the National Natural Science Foundation of China (Grant Nos. 12605177 and 12335005), the PI Research Fund of Henan Normal University (Grant No. 5101029470335), and the Natural Science Foundation of Henan Province (Grant Nos. 262300421233 and 27AQ140008).
The numerical calculations in this work were carried out on the High Performance Computing Platform at the Center for Theoretical Physics, Henan Normal University.

\bibliographystyle{elsarticle-num}
\bibliography{ref}

\end{document}